\documentclass{article} 
\usepackage[hmarginratio=1:1,vmarginratio=1:1,margin=2cm,top=2cm]{geometry}
\usepackage[utf8]{inputenc} 
\usepackage[T1]{fontenc} 
\usepackage{textcomp} 
\usepackage{graphicx} 
\usepackage{subcaption} 
\usepackage{amssymb} 
\usepackage{float} 
\usepackage{multirow} 

\usepackage[style=apa]{biblatex}
\usepackage{caption} 
\usepackage{svg} 
\usepackage{amsmath}
\usepackage{makecell} 
\usepackage{textcomp} 
\usepackage{setspace} 
\usepackage{lineno} 
\usepackage{hyperref} 
\usepackage{orcidlink}
\usepackage{authblk} 
\usepackage{tabto} 
\NumTabs{8} 

\usepackage{xcolor}

\title{Linea alba 3D morphometric variability by CT scan exploration}

\author[1,2]{Pierre Gueroult \orcidlink{0000-0003-3607-1081}}
\author[1]{Victoria Joppin \orcidlink{0000-0003-2789-6647}}
\author[3,4]{Kathia Chaumoitre \orcidlink{0000-0001-9685-5377}}
\author[3]{Mathieu Di Bisceglie \orcidlink{0000-0002-9530-2006}}
\author[1]{Catherine Masson \orcidlink{0000-0003-3578-9067}}
\author[1,5]{Thierry Bege \orcidlink{0000-0002-0775-3035}}

\affil[1]{Univ Gustave Eiffel, Aix-Marseille Univ, LBA, F-13016 Marseille, France}
\affil[2]{Service de chirurgie viscérale et endocrinienne, Angers, University Hospital, Rue Larrey, 49933, CEDEX 9 Angers, France}
\affil[3]{Department of Medical Imaging, Aix Marseille Univ, North Hospital, APHM, Marseille, France}
\affil[4]{Anthropologie Biologique UMR 7268ADES, Aix Marseille Univ, Marseille, France}
\affil[5]{Department of General Surgery, Aix Marseille Univ, North Hospital, APHM, Marseille, France}

\begin{document}

\maketitle

\noindent \textbf{Corresponding author}

Pierre Gueroult \orcidlink{0000-0003-3607-1081} \tab gueroult.pierre@gmail.com

\tableofcontents

\vspace{1cm}

\noindent \textbf{Keywords}

\noindent Linea alba; CT scan; Anatomy; 3D reconstruction; Sagitta

\vspace{0.5cm}

\noindent \textbf{Abbreviations}

\noindent IRD: inter-rectus distance; MPR: multiplanar reconstruction; \textit{r}: Pearson correlation coefficient

\newpage

\section*{Abstract}

\noindent \textit{Purpose}: The width of the Linea alba, which is often gauged by inter-rectus distance, is a key risk factor for incisional hernia and recurrence. Previous studies provided limited descriptions with no consideration for width, location variability, or curvature. We aimed to offer a comprehensive 3D anatomical analysis of the Linea alba, emphasizing its variations across diverse demographics.

\vspace{0.2cm}

\noindent \textit{Methods}: Using open source software, 2D sagittal plane and 3D reconstructions were performed on 117 patients’ CT scans. Linea alba length, curvature assessed by the sagitta (the longest perpendicular segment between xipho-pubic line and the Linea alba), and continuous width along the height were measured.

\vspace{0.2cm}

\noindent \textit{Results}: The Linea alba had a rhombus shape, with a maximum width at the umbilicus of 4.4 ± 1.9 cm and a larger width above the umbilicus than below. Its length was 37.5 ± 3.6 cm, which increased with body mass index (BMI) (\textit{p}<0.001), and was shorter in women (\textit{p}<0.001). The sagitta was 2.6 ± 2.2 cm, three times higher in the obese group (\textit{p}<0.001), majorated with age (\textit{p}=0.009), but was independent of gender (\textit{p}=0.212). Linea alba width increased with both age and BMI (\textit{p}<0.001, \textit{p}=0.002), being notably wider in women halfway between the umbilicus and pubis (\textit{p}=0.007).

\vspace{0.2cm}

\noindent \textit{Conclusion}: This study provides an exhaustive 3D description of Linea Alba’s anatomical variability, presenting new considerations for curvature. This method provides a patient-specific anatomy description of the Linea alba. Further studies are needed to determine whether 3D reconstruction correlates with pathologies, such as hernias and diastasis recti.

\section{Introduction}

\noindent The abdominal wall is a complex, multilayer structure composed of muscles, aponeurosis, fat, and skin. It provides essential functions such as trunk mobility, abdominal pressure regulation, breathing, and organ protection. The Linea alba comprises the entanglement of anterior and posterior rectus sheath \parencite{goldbergNormalRadiographicAnatomy2023,axerCollagenFibersLinea2001a}, with a three-layer structure made of collagen fibers of different orientations \parencite{axerCollagenFibersLinea2001a} and presenting with anisotropic properties \parencite{astrucCharacterizationAnisotropicMechanical2018,levillainContributionCollagenElastin2016,cooneyUniaxialBiaxialTensile2016}. It plays an important role in stabilizing the abdominal wall complex. Its dysfunction leads to parietal pathologies such as hernias and diastasis recti, or participates in syndromes such as postpartum abdominal wall insufficiency (PPAWI) \parencite{smietanskiPostpartumAbdominalWall2022}.

\vspace{0.5cm}

\noindent The usual anatomic treatise descriptions of the Linea alba are not very detailed \parencite{rouviereAnatomieHumaineDescriptive2002,kaminaAnatomieCliniqueTome2014,netterAtlasNetterDanatomie2023}, and few authors have made more precise descriptions based on dissection \parencite{rathAbdominalLineaAlba1996,testutTraiteDanatomieHumaine1948}. Modern postprocessing techniques using CT scans, such as multiplanar reformation, three-dimensional (3D) visualization, and organ delineation, make it possible to study anatomy \textit{in vivo} \parencite{kirchgeorgIncreasingSpiralCT1998,maherTechniquesClinicalApplications2004}. Several authors have described the Linea alba from a radiological point of view \parencite{goldbergNormalRadiographicAnatomy2023,kaufmannNormalWidthLinea2022,jourdanAbdominalWallMorphometric2020,fredonCorrelationsRectusAbdominis2021}. It has been shown to be enlarged in obese people, during pregnancy and postpartum, at higher ages, and with diabetes mellitus \parencite{jourdanAbdominalWallMorphometric2020,cavalliPrevalenceRiskFactors2021,motaNormalWidthInterrecti2018,yuanPrevalenceRiskFactors2021}. However, these studies have been limited to axial planes, with no consideration of their sagittal length and 3D shape. Furthermore, axial studies are limited to selected slices, with no consideration for the entire anatomic profile. 

\vspace{0.5cm}

\noindent Most of the literature studying the Linea alba has been driven by the problem of defining diastasis recti \parencite{kaufmannNormalWidthLinea2022,cavalliPrevalenceRiskFactors2021,motaNormalWidthInterrecti2018,reinpoldClassificationRectusDiastasis2019,motaPrevalenceRiskFactors2015}. The European Hernia Society recommendations define it as an inter-rectus distance (IRD) exceeding 2 cm that should be measured 3 cm above the umbilicus \parencite{hernandez-granadosEuropeanHerniaSociety2021}. With this definition, 35.6-57$\%$ of the population should be considered as presenting a diastasis recti \parencite{kaufmannNormalWidthLinea2022,ugurluPrevalenceRectusDiastasis2023}. Diastasis recti is often associated with umbilical, small midline, and groin hernia \parencite{ugurluPrevalenceRectusDiastasis2023,kohlerSuturedRepairPrimary2015,bellidoluqueTotallyEndoscopicSurgery2015,nishiharaComorbidRectusAbdominis2021}. IRD is also a risk factor for recurrence after umbilical and midline hernia sutured repair \parencite{kohlerSuturedRepairPrimary2015,bellidoluqueTotallyEndoscopicSurgery2015}. The appropriate treatment in patients presenting a hernia associated with diastasis recti is still debated in the surgical field. A recent meta-analysis concluded that both open and laparoscopic approaches are effective, as evidenced by very low recurrence rates \parencite{elhawaryClosingGapEvidencebased2021}.

\vspace{0.5cm}

\noindent The width of the Linea alba depends on the level of its measurement \parencite{rathAbdominalLineaAlba1996,kaufmannNormalWidthLinea2022}, but the assessments that have been carried out to date are \guillemotleft discontinuous\guillemotright{} because of the limited number of measurements at different levels of the Linea alba. The variation of its width throughout the height and its curvature has never been considered in anatomy or in rectus diastasis repair. The IRD is not a constant distance over the entire height of the wall, which should be taken into account.

\vspace{0.5cm}

\noindent We aimed to provide a complete description of the Linea alba’s width, shape, and curvature and to determine which factors influenced its parameters.

\section{Materials and Methods}

\subsection{Cohort’s selection}

\noindent We conducted a retrospective study on a cohort of patients that was previously designed to describe abdominal wall axial morphometrics \parencite{jourdanAbdominalWallMorphometric2020}. The inclusion criteria of the original cohort were as follows: adult patients over 18 years or older performing a CT scan for renal colic from October 2017 to April 2019 at the medical imaging department of the North Hospital, Marseille (France). The exclusion criteria of the original cohort were visible abdominal wall tissue disorder and the presence of a significant medical record (cancer, immunodeficiency, etc.).

\vspace{0.5cm}

\noindent A total of 841 patients met the criteria, and 120 CTs were randomly chosen to make four comparable groups of age (18–30 years, 31–45 years, 46–60 years, and older than 60 years, n = 30/group, n = 15 women in each group) , gender (n = 60/gender), and body mass index (BMI) (under and normal weight [BMI$\leq$25, n = 52], overweight [BMI 25–30, n = 41], and obese [BMI$\geq$30, n = 27]).

\subsection{CT scans}

\noindent CTs were performed on a Revolution HD GSI (GE Healthcare, USA) in a supine position. Patients were told to hold their breath after maximal inhalation during acquisition. No contrast agent was injected. Exposure resulted in a dose of radiation of 120 kV, and the exposure time (mAS) and scan field of view varied according to the patient’s weight. Slice thickness was 1.25 mm, and pitch was 1.375. The matrix size of the images was 512 × 512 pixels.

\subsection{Output metrics from CT scans}

\noindent CT scan measurements were performed by the same operator (PG, medical doctor) using Osirix Lite imaging software (v.13.0.2 Pixmeo SARL, Switzerland). Afterward, 3D multiplanar reconstruction (MPR) median sagittal reconstructions were systematically performed. Reconstruction of the whole Linea alba was not possible in three patients who only had upper abdomen images; those patients were discarded from the analysis. 

\vspace{0.5cm}

\begin{figure}[H]
\centering
\captionsetup{justification=centering, format=plain}
\includegraphics[width=0.6\textwidth]{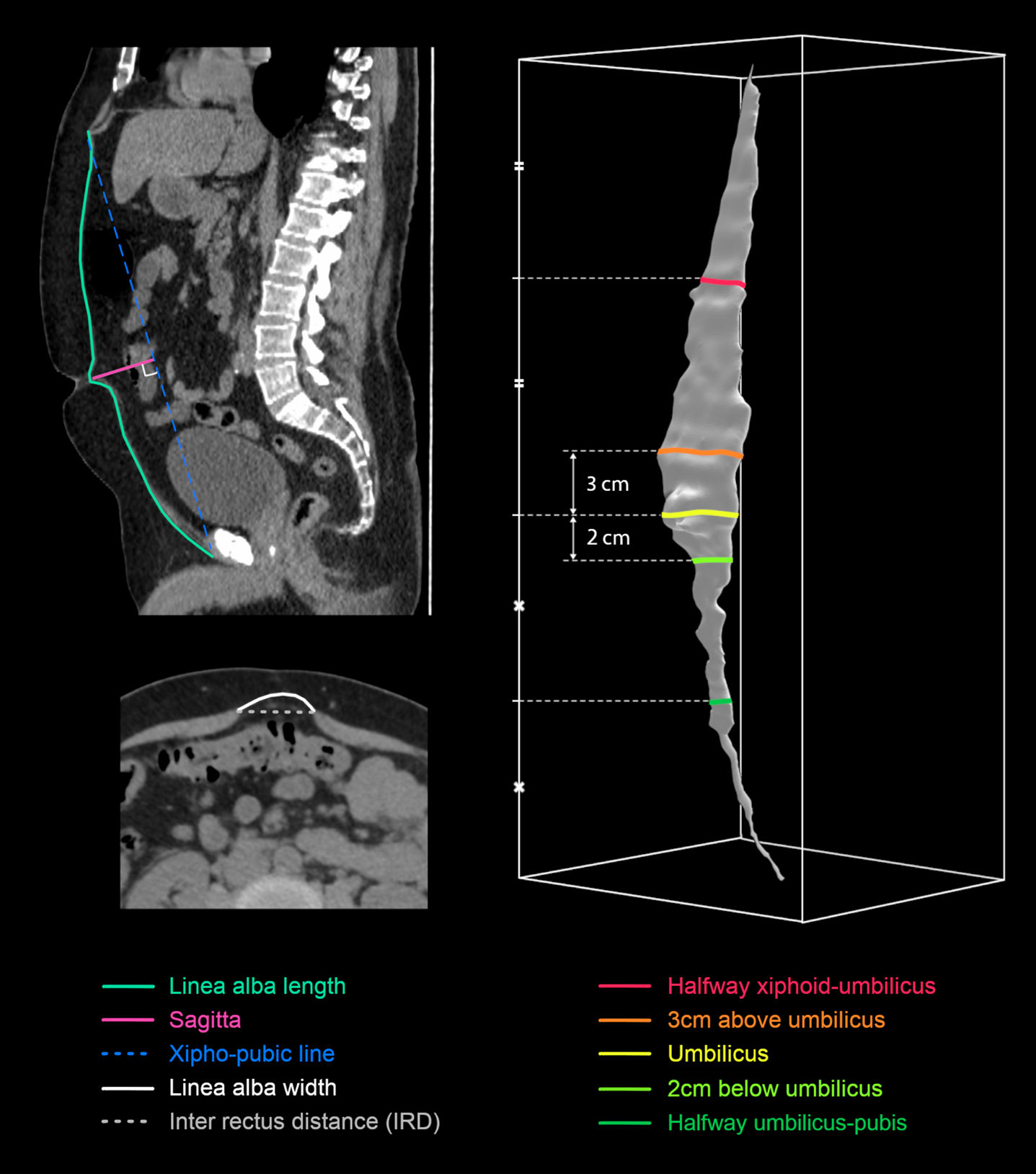}
\caption[Linea alba measurements]{Sagittal, axial and 3D measurements performed on the Linea alba: Linea alba length (light blue), sagitta (pink), Linea alba width (white), Inter rectus distance (grey), width halfway xiphoid-umbilicus (red), width 3 cm above umbilicus (orange), width at umbilicus (yellow), width 2 cm below umbilicus (light green), width halfway umbilicus-pubis (green)}
\label{fig:M_and_M_linea_alba_CT_measurements}
\end{figure}

\noindent Measurements were taken to quantify the length, curvature, and width of the Linea alba. These measurements are shown in \autoref{fig:M_and_M_linea_alba_CT_measurements}.

\noindent The Linea alba length was defined as the distance between its insertion on the xiphoid process and on the pubic bone along the Linea alba. Curvature was described with the sagitta. In geometry, the sagitta of a circular arc is the distance from the center of the arc to the center of its base. Here, the base corresponded to the xipho-pubic line, and the sagitta was defined as the longest perpendicular segment between the xipho-pubic line and the Linea alba. All presented measures were rounded to millimeters.

\vspace{0.5cm}

\noindent A 3D reconstruction of the Linea alba was made for every patient on the open source 3D Slicer software (Version 4.13.0). Delineations of the Linea alba were performed semiautomatically using the 2 mm paintbrush segmentation tool. The Linea alba was delineated manually on approximately 25$\%$ of axial and sagittal slices. The delineation was then propagated using the \guillemotleft fill between slice\guillemotright{} tool. After propagation, the remaining holes were filled manually. Final 3D rendering is presented with a smoothing factor of 0.5. 

\vspace{0.5cm}

\noindent The width of the Linea alba was calculated from the 3D axial plane geometry along its entire height using a specially designed Matlab algorithm (MathWorks ink, Version R2023a). The IRD distance corresponded to the shortest segment between the lateral edges of the Linea alba. The IRD is measured as a straight line, whereas the length of the Linea alba, as measured in this study, is a curve. Maximal and average Linea alba widths were calculated. The width was measured in specific landmarks of Linea alba (from top to bottom): halfway xiphoid–umbilicus, 3 cm above umbilicus, at umbilicus, 2 cm below umbilicus, and halfway umbilicus–pubis. These landmarks corresponded to those used to define diastasis recti and to those usually found in the literature \parencite{kaufmannNormalWidthLinea2022,motaNormalWidthInterrecti2018,hernandez-granadosEuropeanHerniaSociety2021}. A relative Linea alba length scale suitable for all patients, regardless of their body size, was established for the location of maximal width and graphical comparison between individuals. The average width of the Linea alba at each level of its standardized height was then plotted on the same axis, group by group, with key anatomical landmarks.

\vspace{0.5cm}

\noindent Morphometric measurements performed at the upper edge of the fourth lumbar vertebrae and previously described in Jourdan \textit{et al.} study \parencite{jourdanAbdominalWallMorphometric2020} were extracted and analyzed in the present cohort: total abdominal area, total visceral area, muscular area, abdominal muscles area, subcutaneous fat area, visceral fat area, subcutaneous fat thickness, and waist circumference.

\subsection{Statistical analysis}

\noindent Statistical analysis was performed using RStudio (Version 2022.7.2.576 RStudio, Inc., Boston, MA). Quantitative variables were expressed as means $\pm$ one standard deviation, minimal values, and maximal values. The normality of the distribution was checked using a Shapiro–Wilk test. The normally distributed variable was Linea alba length. Non-normally distributed variables were sagitta and all Linea alba width measurements. When required, equality of variance was checked before the statistical analysis. Differences between sex groups were tested using the Student t-test for normally distributed variables and the Mann–Whitney for non-normally distributed variables. The differences between age groups and BMI groups were tested using an analysis of variance (ANOVA) for normally distributed variables and a Kruskal–Wallis test when not normally distributed. The normality of residuals was checked using a Shapiro–Wilk test. Alpha risk was 5$\%$, two-tailed. Here, \textit{p}-values under 0.025 were considered to be statistically significant. The correlation between variables was calculated using Pearson’s correlation coefficient \textit{r}. Data are presented using a correlation matrix plot coding Pearson’s correlation coefficient with a gradation of colors. In this figure, BMI and age are treated as continuous variables. For the reader’s convenience, the raw correlation coefficients have not been presented in the corresponding figure. The most representative patient of each subgroup was the one with the shortest Euclidean distance to the mean of all variables in each subgroup.

\section{Results}

\subsection{Demographic data}

\noindent The final cohort was made up of 117 patients: 58 women and 59 men. The mean age was 45.5 $\pm$ 16.4 years old, and the mean BMI was 26.1 $\pm$ 4.9 $kg/m^2$.

\subsection{Linea alba measurements}
\noindent The average length of the Linea alba in the sagittal plane was 37.5 $\pm$ 3.6 cm (range [30.5/47.7 cm]). The mean sagitta was 2.6 $\pm$ 2.2 cm (range [-3.4/8.1 cm]). Some patients (n=9) had a negative sagitta because of an inward curved Linea alba. 

\begin{figure}[H]
\centering
\includegraphics[width=0.9\textwidth]{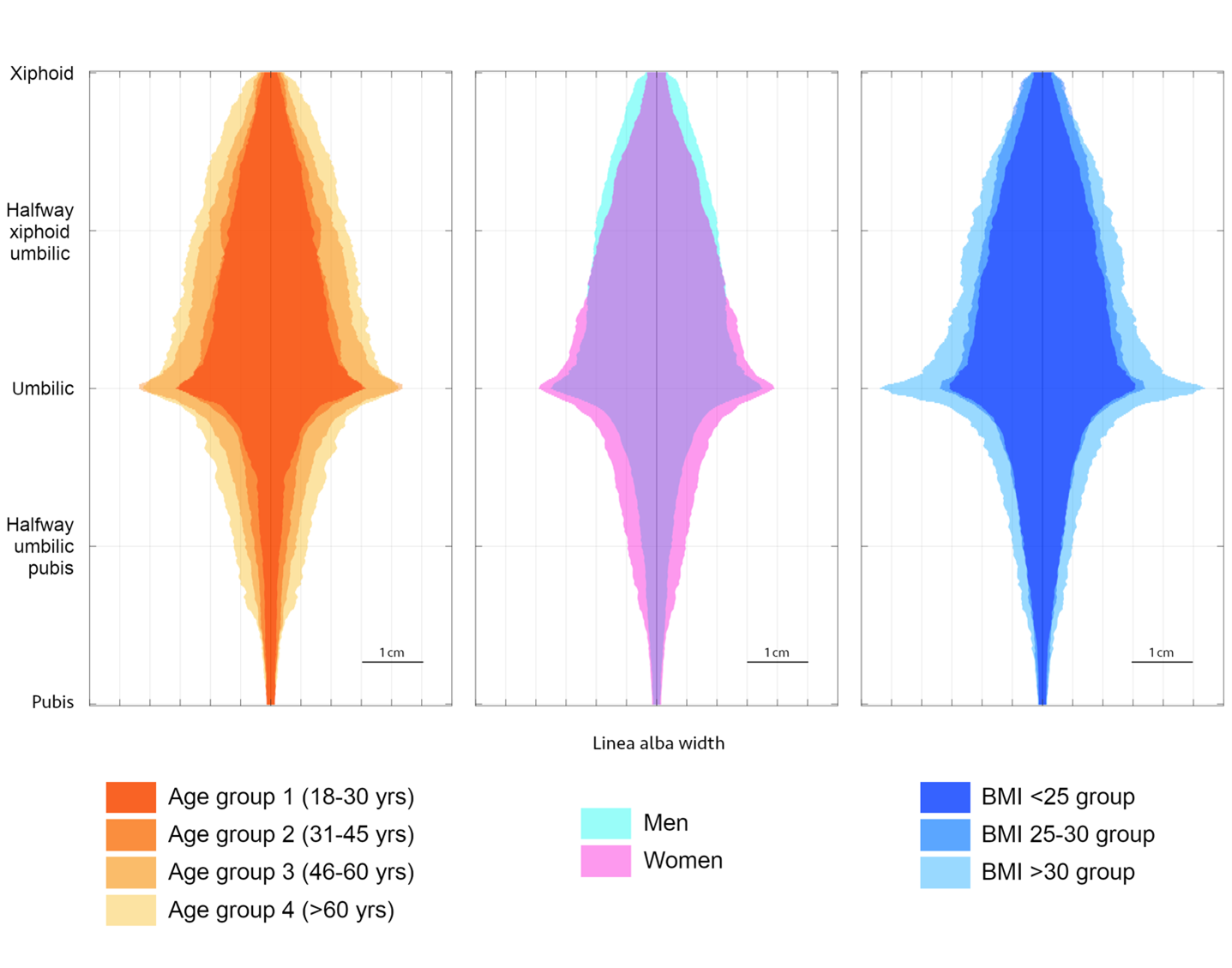}
\captionsetup{justification=centering, format=plain}
\caption[Linea alba width according to age, gender, and BMI groups]{Linea alba width calculated on 3D reconstruction according to age, gender, and BMI groups. Linea alba length has been standardized between individuals to allow for a comparison of key landmarks}
\label{fig:LB_3D_reconstruction}
\end{figure}

\begin{figure}[H]
\centering
\captionsetup{justification=centering, format=plain}
\includegraphics[width=0.9\textwidth]{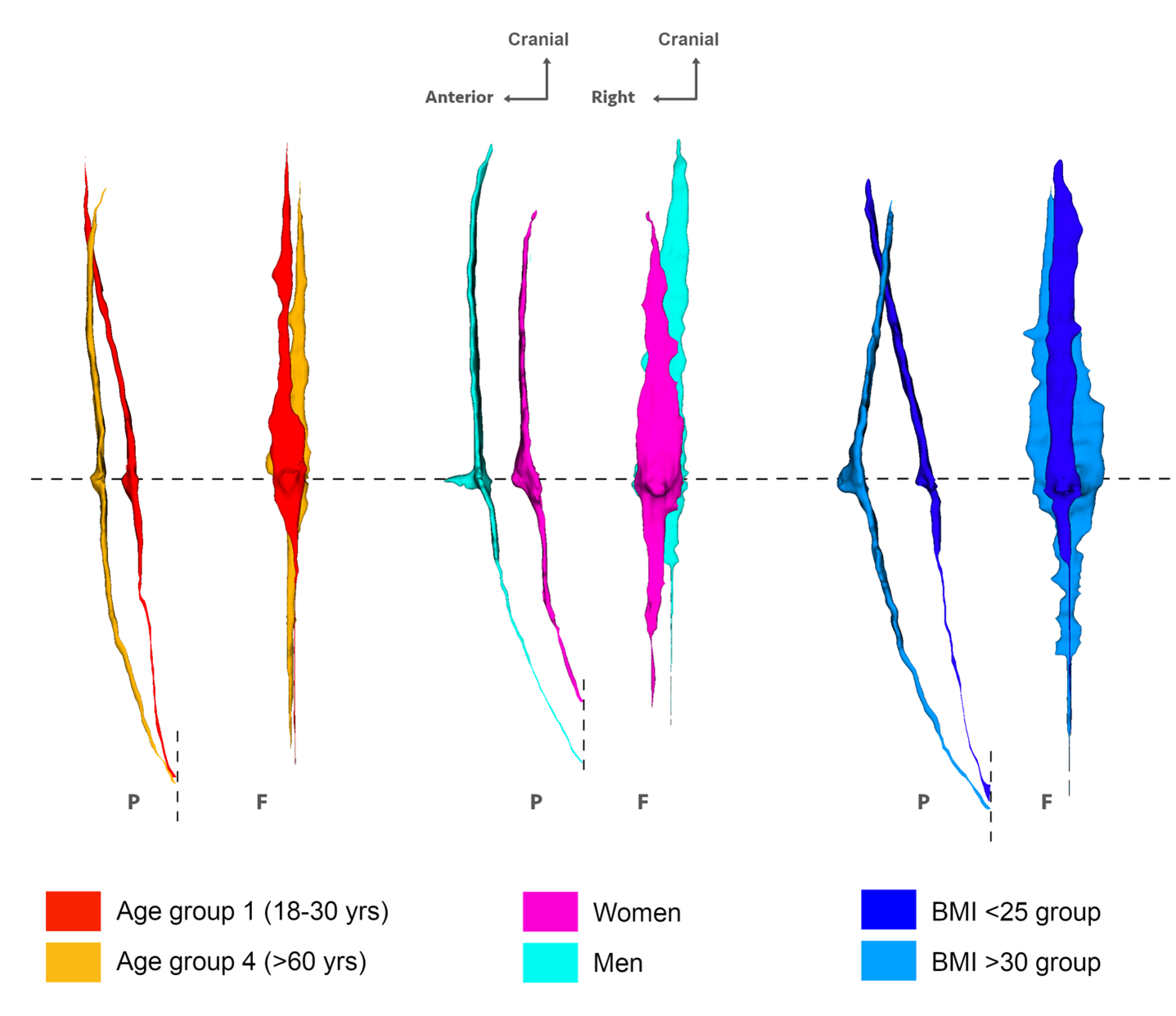}
\caption[Linea alba of the most representative patient]{3D reconstruction of the Linea alba in the most representative patient in each group. The face (F) and profile (P) of the reconstructions are presented. Linea alba are aligned on the umbilicus and pubic insertion levels. The most representative patients are those with the shortest Euclidean distance to the mean of all variables in each subgroup. Reconstructions has been generated with a 3D slicer}
\label{fig:LB_3D_reconstruction_most_representative_patient}
\end{figure}

\noindent Linea alba width measurement showed a tendency to be rhombus shaped (\autoref{fig:M_and_M_linea_alba_CT_measurements}, \autoref{fig:LB_3D_reconstruction}, \autoref{fig:LB_3D_reconstruction_most_representative_patient}). The average maximal width was 4.4 $\pm$ 1.9 (range [1.4/11.4 cm]) an was located at the umbilicus level. Linea alba widths at specific landmarks are presented in \autoref{tab:demographic_data}.

\begin{table}[H]
\renewcommand{\arraystretch}{1.5}
\begin{center}
\resizebox{\columnwidth}{!}{
\begin{tabular}{|c|c|c|}
\hline
\textbf{}                                       & \textbf{Mean $\pm$ Standard deviation} & \textbf{Range [min/max]}  \\ \hline
\multicolumn{3}{|c|}{\textbf{Demographic data}}                                                                     \\ \hline
\textbf{Age} ($years$)                             & 45.5 $\pm$ 16.4                       & [18.0/86.0]         \\ \hline
\textbf{Size} ($m$)                               & 1.69 $\pm$ 0.10                       & [1.45/1.92]         \\ \hline
\textbf{BMI} ($kg/m^2$)                        & 26.1 $\pm$ 4.9                        & [15.6/43.3]         \\ \hline
\multicolumn{3}{|c|}{\textbf{Linea alba measurements} ($cm$)}                                          \\ \hline
\textbf{Sagitta}                               & 2.6 $\pm$ 2.2                         & [-3.4/8.1]          \\ \hline
\textbf{Length}                            & 37.5 $\pm$ 3.6                        & [30.5/47.7]         \\ \hline
\multicolumn{3}{|c|}{Width ($cm$)}                                                                       \\ \hline
\textbf{Halfway xiphoid-umbilicus}              & 1.9 $\pm$ 1.2                         & [0.2/8.1]           \\ \hline
\textbf{3 cm above the umbilicus}               & 2.6 $\pm$ 1.4                         & [0.6/9.6]           \\ \hline
\textbf{At the umbilicus}                       & 3.7 $\pm$ 1.9                         & [0.9/11]           \\ \hline
\textbf{2 cm below the umbilicus}               & 1.7 $\pm$ 1.2                         & [0.4/9.6]           \\ \hline
\textbf{Halfway umbilicus-pubis}                & 0.7 $\pm$ 0.9                         & [0.1/5.6]           \\ \hline
\textbf{Maximal width}                     & 4.4 $\pm$ 1.9                         & [1.4/11.4]          \\ \hline
\textbf{Maximal inter-rectus distance (IRD)}              & 3.1 $\pm$ 1.4                         & [1.1/9.4]           \\ \hline
\multicolumn{3}{|c|}{\textbf{Other abdominal measurements}}                                                   \\ \hline
\multicolumn{3}{|c|}{Length ($cm$)}                                                                    \\ \hline
\textbf{Subcutaneous fat thickness}        & 2.3 $\pm$ 0.9                         & [0.4/4.4]           \\ \hline
\textbf{Waist circumference}               & 95.2 $\pm$ 13.3                       & [68.2/135.3]        \\ \hline
\multicolumn{3}{|c|}{Areas ($cm^2$)}                                                               \\ \hline
\textbf{Total abdominal area}             & 678.4 $\pm$ 199.0                     & [309.1/1318.4]      \\ \hline
\textbf{Total visceral area}              & 270.0 $\pm$ 102.9                     & [87.0/626.0]        \\ \hline
\textbf{Muscular area}                    & 133.2 $\pm$ 32.4                      & [74.2/215.3]        \\ \hline
\textbf{Abdominal muscles area}           & 55.6 $\pm$ 16.6                       & [24.4/102.1]        \\ \hline
\textbf{Subcutaneous fat area}            & 249.3 $\pm$ 125.9                     & [44.9/752.7]        \\ \hline
\textbf{Visceral fat area}                & 147.0 $\pm$ 94.0                      & [17.2/449.0]        \\ \hline
\end{tabular}
}
\end{center}
\captionsetup{justification=centering, format=plain}
\caption[Demographic data and linea alba measurements]{Demographic data, Linea alba measurements, and other abdominal measurements for the whole cohort (n=117) \\ Data are presented as mean $\pm$ standard deviation}
\label{tab:demographic_data}
\end{table}

\subsection{Influence of age}

\noindent Age had a significant impact on all abdominal metrics, except Linea alba length (\autoref{tab:influence_age}, \autoref{fig:LB_3D_reconstruction} and \autoref{fig:LB_3D_reconstruction_most_representative_patient}). The Linea alba widened with age in all landmarks, regardless of the size (\textit{p}<0.001, \textit{p}=0.002), but did not lengthen (\textit{p}=0.252).

\noindent Its curvature, as reflected by the sagitta metric, increased significantly toward the front with age (\textit{p}=0.009). 

\begin{table}[H]
\renewcommand{\arraystretch}{2}
\centering
\begin{center}
\resizebox{\columnwidth}{!}{%
\begin{tabular}{|c|c|c|c|c|c|}
\hline
\textbf{Linea alba measurements*} ($cm$) & \makecell{\textbf{Age group 1} \\ (18–30 years) n=29} & \makecell{\textbf{Age group 2} \\ (31–45 years) n=30} & \makecell{\textbf{Age group 3} \\ (46–60 years) n=30} & \makecell{\textbf{Age group 4} \\ (>60 years) n=28} & \textbf{\textit{p}-value}                          \\ \hline
\textbf{Sagitta}                 & 2.2 $\pm$ 2.2  & 2.0 $\pm$ 2.0  & 3.1 $\pm$ 2.1  & 3.3 $\pm$ 2.2 & 0.009 **       \\ \hline
\textbf{Length}                  & 37.7 $\pm$ 3.1 & 38.0 $\pm$ 4.2 & 38.0 $\pm$ 3.5 & 36.4 $\pm$ 3.2 & 0.252         \\ \hline
\multicolumn{6}{|c|}{Width}                                                                                          \\ \hline
\textbf{Halfway xiphoid-umbilicus}      & 1.4 $\pm$ 0.7  & 1.7 $\pm$ 1.0  & 2.2 $\pm$ 1.1  & 2.5 $\pm$ 1.5 & <0.001 **      \\ \hline
\textbf{3 cm above the umbilicus}       & 2.1 $\pm$ 0.9  & 2.2 $\pm$ 1.0  & 3.0 $\pm$ 1.2  & 3.1 $\pm$ 2.0 & 0.002 **       \\ \hline
\textbf{At the umbilicus}               & 3.2 $\pm$ 1.7  & 3.1 $\pm$ 1.6  & 4.4 $\pm$ 1.7  & 4.2 $\pm$ 2.1 & 0.002 **       \\ \hline
\textbf{2 cm below the umbilicus}       & 1.1 $\pm$ 0.5  & 1.3 $\pm$ 0.6  & 1.9 $\pm$ 1.2  & 2.3 $\pm$ 1.8 & <0.001 **      \\ \hline
\textbf{Halfway umbilicus-pubis}        & 0.3 $\pm$ 0.3  & 0.5 $\pm$ 0.5  & 0.9 $\pm$ 0.9  & 1.2 $\pm$ 1.2 & <0.001 **      \\ \hline
\textbf{Maximal width}                  & 3.6 $\pm$ 1.6  & 3.7 $\pm$ 1.5  & 5.1 $\pm$ 1.8  & 5.4 $\pm$ 2.2 & <0.001 **      \\ \hline
\textbf{Maximal inter-rectus distance (IRD)}           & 2.4 $\pm$ 0.9  & 2.6 $\pm$ 0.9  & 3.5 $\pm$ 1.2  & 4.1 $\pm$ 1.8 & <0.001 **      \\ \hline
\end{tabular}
}
\end{center}
\captionsetup{justification=centering, format=plain}
\caption[Influence of age on linea alba measurements]{Influence of age on Linea alba measurements \\ *Data are presented as mean $\pm$ standard deviation \\ **Indicate statistically significant \textit{p}-values (\textit{p}<0.025)}
\label{tab:influence_age}
\end{table}

\subsection{Influence of gender}

\noindent Gender had a significant impact on several measurements (\autoref{tab:influence_gender}, \autoref{fig:LB_3D_reconstruction} and \autoref{fig:LB_3D_reconstruction_most_representative_patient}).

\noindent Women had a significantly wider Linea alba at the halfway umbilicus–pubis landmark (\textit{p}=0.007). No statistical difference between men and women was found for width at other landmarks. Linea alba length (\textit{p}<0.001) was significantly lower in women. The curvature reflected by the sagitta metric was not statistically different between men and women (\textit{p}=0.212). This suggests the stability of this measurement between genders.

\noindent Visually, men tended to have a wider Linea alba toward the top but without statistical significance.

\begin{table}[H]
\renewcommand{\arraystretch}{2}
\centering
\begin{center}
\begin{tabular}{|c|c|c|c|}
\hline
\textbf{Linea alba measurements*} ($cm$)   & \makecell{\textbf{Men} \\ n=59}   & \makecell{\textbf{Women} \\ n=58} & \textbf{\textit{p}-value}   \\ \hline
\textbf{Sagitta}               & 2.9 $\pm$ 2.0         & 2.4 $\pm$ 2.3         & 0.212                         \\ \hline
\textbf{Length}                & 38.6 $\pm$ 3.8        & 36.4 $\pm$ 2.9        & <0.001 **                      \\ \hline
\multicolumn{4}{|c|}{Width}                                                                       \\ \hline
\textbf{Halfway xiphoid–umbilicus}  & 2.0 $\pm$ 1.1         & 1.8 $\pm$ 1.2         & 0.184                         \\ \hline
\textbf{3 cm above the umbilicus}   & 2.5 $\pm$ 1.3         & 2.7 $\pm$ 1.5         & 0.281                         \\ \hline
\textbf{At the umbilicus}           & 3.5 $\pm$ 1.6         & 3.9 $\pm$ 2.1         & 0.417                         \\ \hline
\textbf{2 cm below the umbilicus}   & 1.4 $\pm$ 0.6         & 2.0 $\pm$ 1.6         & 0.058                         \\ \hline
\textbf{Halfway umbilicus–pubis}    & 0.5 $\pm$ 0.5         & 1.0 $\pm$ 1.1         & 0.007 **                      \\ \hline
\textbf{Maximal width}         & 4.2 $\pm$ 1.6         & 4.6 $\pm$ 2.2         & 0.515                         \\ \hline
\textbf{Maximal inter-rectus distance (IRD)}  & 3.0 $\pm$ 1.2         & 3.3 $\pm$ 1.6         & 0.689                         \\ \hline
\end{tabular}
\end{center}
\captionsetup{justification=centering, format=plain}
\caption[Influence of gender on linea alba measurements]{Influence of gender on Linea alba measurements \\ *Data are presented as mean $\pm$ standard deviation \\ **Indicate statistically significant \textit{p}-values (\textit{p}<0.025)}
\label{tab:influence_gender}
\end{table}

\subsection{Influence of BMI}

\noindent BMI had a huge impact on abdominal metrics (\autoref{tab:influence_BMI}, \autoref{fig:LB_3D_reconstruction} and \autoref{fig:LB_3D_reconstruction_most_representative_patient}).

\noindent Linea alba width was greater at the level of the umbilicus and around the umbilicus in obese people (\textit{p}<0.001), whereas it was not greater from halfway xiphoid–umbilicus and halfway umbilicus–pubis (\textit{p}=0.142 and \textit{p}=0.071). Linea alba length was significantly greater with BMI increase (\textit{p}<0.001).

\noindent The curvature, as reflected by the sagitta metric, was statistically higher with BMI increase (\textit{p}<0.001). The sagitta was even three times higher in the BMI>30 group than in the BMI<25 group (4.8 $\pm$ 2.0 cm versus 1.6 $\pm$ 1.7 cm).

\noindent This indicates an overall widening of the Linea alba along its entire length, with a maximum in the periumbilical area in obese people.

\begin{table}[H]
\renewcommand{\arraystretch}{2}
\centering
\begin{center}
\begin{tabular}{|c|c|c|c|c|}
\hline
\textbf{Linea alba measurements*} ($cm$)   & \makecell{\textbf{BMI<25} \\ n=51}    & \makecell{\textbf{BMI 25–30} \\ n=40}     & \makecell{\textbf{BMI>30} \\ n=26}      & \textbf{\textit{p}-value}  \\ \hline
\textbf{Sagitta}               & 1.6 $\pm$ 1.7             & 2.5 $\pm$ 1.8                 & 4.8 $\pm$ 2.0               & <0.001 **     \\ \hline
\textbf{Length}                & 35.9 $\pm$ 2.9            & 37.6 $\pm$ 3.3                & 40.7 $\pm$ 3.0              & <0.001 **     \\ \hline
\multicolumn{5}{|c|}{Width}                                                                                                            \\ \hline
\textbf{Halfway xiphoid–umbilicus}  & 1.7 $\pm$ 1.0             & 1.9 $\pm$ 1.0                 & 2.4 $\pm$ 1.7               & 0.142         \\ \hline
\textbf{3 cm above the umbilicus}   & 2.3 $\pm$ 1.3             & 2.5 $\pm$ 1.4                 & 3.3 $\pm$ 1.3               & <0.001 **     \\ \hline
\textbf{At the umbilicus}           & 3.1 $\pm$ 1.8             & 3.4 $\pm$ 1.2                 & 5.4 $\pm$ 1.9               & <0.001 **     \\ \hline
\textbf{2 cm below the umbilicus}   & 1.5 $\pm$ 1.4             & 1.5 $\pm$ 0.7                 & 2.3 $\pm$ 1.3               & <0.001 **     \\ \hline
\textbf{Halfway umbilicus–pubis}    & 0.6 $\pm$ 1.0             & 0.7 $\pm$ 0.6                 & 1.0 $\pm$ 0.9               & 0.071         \\ \hline
\textbf{Maximal width}              & 3.7 $\pm$ 1.8             & 4.2 $\pm$ 1.5                 & 6.3 $\pm$ 1.8               & <0.001 **     \\ \hline
\textbf{Maximal inter-rectus distance (IRD)}       & 2.7 $\pm$ 1.3             & 3.1 $\pm$ 1.1                 & 4.1 $\pm$ 1.5               & <0.001 **     \\ \hline
\end{tabular}
\end{center}
\captionsetup{justification=centering, format=plain}
\caption[Influence of BMI on linea alba measurements]{Influence of BMI on Linea alba measurements \\ *Data are presented as mean $\pm$ standard deviation \\ **Indicate statistically significant \textit{p}-values (\textit{p}<0.025)}
\label{tab:influence_BMI}
\end{table}

\subsection{Correlations}

\begin{figure}[H]
\centering
\captionsetup{justification=centering, format=plain}
\includegraphics[width=0.9\textwidth]{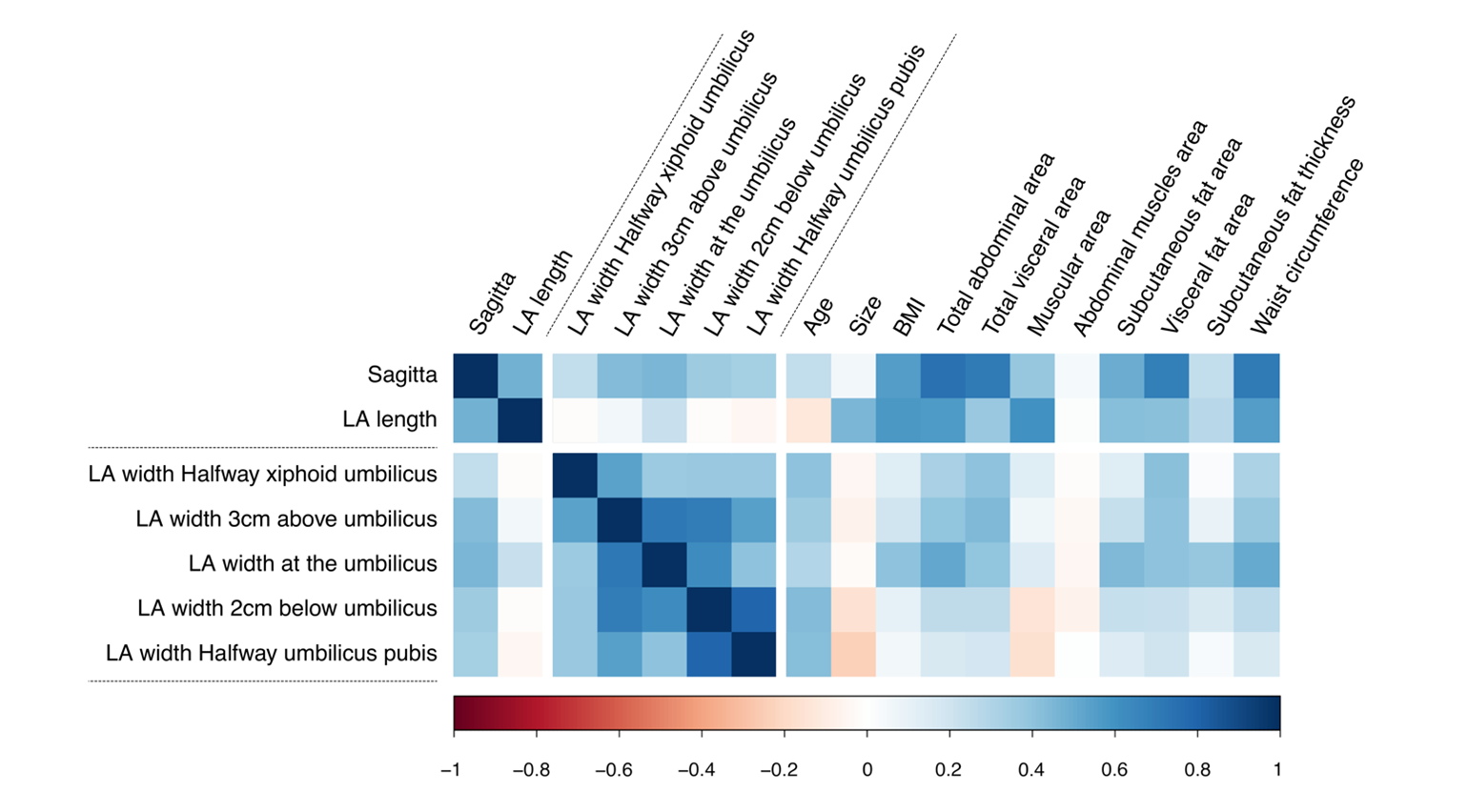}
\caption[Graphical representation of Pearson’s correlation coefficient \textit{r}]{Graphical representation of Pearson’s correlation coefficient \textit{r} between sagittal measurements, axial measurements, and other abdominal measurements. Positive Pearson’s correlation coefficient \textit{r} are shown in shades of blue and negative Pearson’s correlation coefficient \textit{r} in shades of red}
\label{fig:PCC}
\end{figure}

\noindent The correlations between sagittal measurements, axial measurements, and other abdominal measurements are presented in \autoref{fig:PCC}.

\noindent Curvature measurements by the mean of sagitta metric were quasi totally independent of size (\textit{r}=0.06) and poorly influenced by age (\textit{r}=0.24), but these measurements were well correlated with mensuration related to obesity such as umbilical–spinal distance (\textit{r}=0.83), total abdominal area (\textit{r}=0.74), total visceral area (\textit{r}=0.70), visceral fat area (\textit{r}=0.69), and waist circumference (\textit{r}=0.71).

\noindent Linea alba length was weakly associated with Linea alba width at the umbilicus (\textit{r}=0.22). Linea alba length was totally independent of Linea alba width in other locations, such as halfway xiphoid–umbilicus (\textit{r}=0.01), 3 cm above umbilicus (\textit{r}=0.05), 2 cm below umbilicus (\textit{r}=-0.01), and halfway umbilicus–pubis (\textit{r}=-0.04). Linea alba length was more strongly correlated with BMI (\textit{r}=0.57) than with the size of the patient (\textit{r}=0.46). 

\noindent The widths of the Linea alba showed similar correlation profiles, regardless of the landmark. However, the width at the umbilical level was slightly better correlated with the other variables: \textit{r}=0.52 with total abdominal area and \textit{r}=0.51 with waist circumference. Linea alba width at the umbilicus was well associated with its width 2 cm below and 3 cm above (\textit{r}=0.62 and \textit{r}=0.72). Overall, the other measurements were only weakly associated with the Linea alba width.

\section{Discussion}

\noindent The present work has provided a continuous measurement of the width, curvature, and 3D shape of the Linea alba over its entire height. The current study has shown that the maximum width is located at the umbilicus level and that the Linea alba widens and curves forward with age and BMI. Our data have confirmed the previous anatomical descriptions in the axial plane of the Linea alba \parencite{rathAbdominalLineaAlba1996,testutTraiteDanatomieHumaine1948,kaufmannNormalWidthLinea2022,liawRelationshipsInterrectiDistance2011,motaImmediateEffectsInterrectus2015} while adding important nuances based on known risk factors for diastasis and hernia surgery. 

\vspace{0.5cm}

\noindent Until now, \textit{in vivo} sagittal measures of the Linea alba have never been described. Numerous authors have previously ruled on Linea alba width, describing its variability on cranio–caudal axis \parencite{axerCollagenFibersLinea2001a,rathAbdominalLineaAlba1996,testutTraiteDanatomieHumaine1948,kaufmannNormalWidthLinea2022,liawRelationshipsInterrectiDistance2011,motaImmediateEffectsInterrectus2015}. We found similar results with a mean Linea alba width of 2.6 cm measured 3 cm above the umbilicus and a mean of 1.7 cm measured 2 cm below the umbilicus. These data were in accordance with the literature, even if the measurement levels were not homogeneous between the authors \parencite{rathAbdominalLineaAlba1996,testutTraiteDanatomieHumaine1948,kaufmannNormalWidthLinea2022,liawRelationshipsInterrectiDistance2011,motaImmediateEffectsInterrectus2015}. Nevertheless, the Linea alba width at the umbilicus was higher in our study (3.7 $\pm$ 1.9 cm) than in the literature. We found a mean maximal width 1.3 cm higher than the mean maximal inter-rectus distance (IRD). This could be explained by the systematic use of IRD in the literature, while we chose to measure the Linea alba. As visible in \autoref{fig:LB_3D_reconstruction_most_representative_patient}, the Linea alba is curved anteriorly by umbilical defect. As a result, the width of the Linea alba is measured on a curved surface, which means that this measurement is systematically greater than the IRD. The IRD is often chosen because this is how the diastasis recti is defined \parencite{hernandez-granadosEuropeanHerniaSociety2021}. Most of the literature available on the Linea alba is clinical, aiming to define a cut-off for diastasis recti diagnosis. We chose to specifically focus on the Linea alba’s shape, with no consideration for the rectus abdominis muscles.

\vspace{0.5cm}

\noindent In our study, the average length of the Linea alba was 37.5 $\pm$ 3.6 cm. This was greater than the length described on cadaver. Rath \textit{et al.} found a mean of 29.11 cm on 40 subjects, while Axer \textit{et al.} described 26.2 $\pm$ 3.64 cm \parencite{axerCollagenFibersLinea2001a,rathAbdominalLineaAlba1996}. This observation could be explained by the specificity of the cadaver: the use of often old subjects, the conservation process responsible for shrinkage, and rigidity could bias the measurement of Linea alba. This underlines the need for \textit{in vivo} studies to characterize abdominal wall structures.

\vspace{0.5cm}

\noindent In the present study, we introduced the concept of measuring the sagitta in the anatomy of the abdominal wall. Sagitta has never been used to characterize the Linea alba \textit{in vivo}. We found this measurement interesting in assessing the anterior deformation of Linea alba. Our data suggest that the sagitta is higher in obese and elderly patients. Moreover, it is not influenced by gender. It is well correlated with obesity-related measures such as total abdominal area (r=0.74), total visceral area (r=0.70), visceral fat area (r=0.69), and waist circumference (r=0.71). This could be an indirect indicator of Linea alba dehiscence or intra-abdominal pressure. In 2012, Naraynsingh \textit{et al.} supported the hypothesis of a \guillemotleft Sick Linea Alba Complex\guillemotright{} influenced by age, BMI, and parity, resulting in a thin and large Linea alba \parencite{naraynsinghStrongLineaAlba2012}. A measure as easy to obtain as the sagitta could be added to this definition.

\vspace{0.5cm}

\noindent A 3D reconstruction of the Linea alba had never been seen in previous anatomical studies. This recent approach based on affordable tools provides a better understanding of real-life \textit{in vivo} anatomy of Linea alba. Reconstruction was based on axial and sagittal delineation before 3D reconstruction. This allowed for better individualization of the Linea alba and the anterior rectus sheaths. Thus, there was no confusion between Linea alba and IRD. This misunderstanding is often found in the literature \parencite{kaufmannNormalWidthLinea2022,liawRelationshipsInterrectiDistance2011,motaImmediateEffectsInterrectus2015}, even if it has been well illustrated by Beer \textit{et al.}’s dissections and in traditional anatomy treatises that the Linea alba is a structure that can be differentiated from the anterior rectus sheath \parencite{kaminaAnatomieCliniqueTome2014,beerNormalWidthLinea2009}. 

\vspace{0.5cm}

\noindent The use of 3D reconstruction and 3D models is becoming increasingly widespread in the field of medical research. Recently, some authors used finite element simulation to simulate wound healing \parencite{karkhanehyousefiPatientspecificComputationalSimulations2023} or the stability of hernia-damaged abdominal wall \parencite{karrechBiomechanicalStabilityHerniadamaged2023}. These models are not intended to describe patient-specific anatomy and are based on images of a single healthy or cadaveric subject. However, the delineation of the Linea alba is precise, allowing the authors to draw observations based on reliable anatomy.

\vspace{0.5cm}

\noindent Our data have suggested a widening of Linea alba with age and BMI, which is in line with the data in the literature \parencite{kaufmannNormalWidthLinea2022,jourdanAbdominalWallMorphometric2020,fredonCorrelationsRectusAbdominis2021}. Gender only influences lower width, with larger Linea alba halfway umbilicus–pubis in women. One explanation could be parity, which is known to be a risk factor for diastasis recti \parencite{smietanskiPostpartumAbdominalWall2022,motaImmediateEffectsInterrectus2015}. However, these data were not available in the initial cohort, so we cannot conclude the influence of this parameter on our observations.

\vspace{0.5cm}

\noindent Most of the articles available on the Linea alba deal with diastasis recti. According to the European Hernia Society guidelines on the management of diastasis recti, this is defined as an IRD greater than 2 cm. The classification is D1 for an IRD between 2 and 3 cm, D2 for an IRD between 3 and 5 cm, and D3 for an IRD of 5 cm or greater \parencite{hernandez-granadosEuropeanHerniaSociety2021}. IRD of 2 cm up to 3 cm at umbilical level is usually accepted to be physiological \parencite{reinpoldClassificationRectusDiastasis2019}. Regarding our findings, the definition of diastasis recti should be rethought and based on Linea alba width charts. Current data available on Linea alba width could make it possible to represent normal curves and corridors according to age, BMI, and parity. These tools, analogous to growth curves, could be useful in everyday abdominal wall and plastic surgeries. The need for precision is even greater in the context of an aging population prone to obesity. It seems that the pathophysiology of umbilical hernia and diastasis recti is part of a continuous process of degeneration. The delineation of the Linea alba almost systematically revealed a depression corresponding to the umbilical orifice that seemed to widen with age and increasing BMI. This makes it all the more difficult to set a radiological cut-off for umbilical hernia, corroborating the \guillemotleft Sick Linea Alba Complex\guillemotright{} theory of Naraynsingh \textit{et al.} \parencite{naraynsinghStrongLineaAlba2012}. We agree with Kaufmann \textit{et al.}’s conclusions, which called for more population-based studies and a critical re-evaluation of the morphological definition of diastasis recti \parencite{kaufmannNormalWidthLinea2022}.

\subsection{Limitations}

\noindent This study has several limitations. Our algorithm, which was designed to measure Linea alba width, was not able to identify it in 16$\%$ of all data measured in the cohort. This mainly corresponds to near pubis infraumbilical Linea alba, where it is too thin to allow the software to find two separate points. We chose to use a linear implementation to correct data, but the results near extremities should be taken with caution. The use of CT has sometimes made it difficult to identify the beginning of the Linea alba, particularly in patients with a cartilaginous xiphoid process, which is difficult to identify with this imaging modality. The thickness of the Linea alba could not be assessed in this study. This is mainly due to the resolution of the propagation tool used to create the 3D shapes which fails when used for thicknesses of less than 2 mm. This limitation forced us to use the 2 mm brush tool, which is sometimes greater than the expected thickness of the Linea Alba \parencite{axerCollagenFibersLinea2001a}. Finally, the static decubitus position required for the CTs means that we cannot extrapolate these data to the standing position. This last remark argues in favor of carrying out dynamic studies in more physiological positions.

\vspace{0.5cm}

\noindent To the best of our knowledge, the present study is the first to deal with an \textit{in vivo} description of the Linea alba from a multiaxial point of view. It provides a new vision of this structure, which is known to be implicated in parietal pathology. Further studies are needed to evaluate the link between sagittal measurements as risk factors for abdominal wall pathologies. The study of the Linea alba from a dynamic and \textit{in vivo} perspective is still not the standard, even though this point has been raised by Rath \textit{et al.} since 1996 \parencite{rathAbdominalLineaAlba1996}. This opens the door to other explorations using dynamic imaging of the abdominal wall in the axial, sagittal, and even possibly frontal planes and 3D reconstruction. We hope that these data will enrich future 3D models to make abdominal wall surgery simulations more reliable.

\section{Conclusion}

\noindent The present study has provided an exhaustive 3D description of the anatomical variability of the Linea alba with new considerations for curvature expressed by the sagitta. This method, based on open source affordable software, provided a patient-specific anatomy description of the Linea alba, which could be automated in the future. Further studies are needed to determine whether 3D reconstruction correlates with pathologies, such as hernias and diastasis recti, which could make it a standard in patient management.

\vspace{1cm}

\noindent \textbf{Author contributions}

\noindent \textbf{PG}: collected the data

\noindent \textbf{PG} and \textbf{VJ}: processed the data and made figures

\noindent \textbf{VJ}: created the Matlab code

\noindent \textbf{PG}: drafted the article

\noindent \textbf{VJ}, \textbf{KC}, \textbf{MDB}, \textbf{CM} and \textbf{TB}: did the revision of the article.

\vspace{0.5cm}

\noindent \textbf{Funding}

\noindent Pierre GUEROULT disclaimed an \guillemotleft Association Française de Chirurgie\guillemotright{} grant and a state research grant. Grants had no influence on scientifical content of the manuscript.

\vspace{0.5cm}

\noindent \textbf{Data availability}

\noindent All data supporting the findings are available within the paper and its Supplementary Information.

\section*{Declarations}

\noindent \textbf{Conflict of interest}

\noindent The authors state that they do not have any conflicts of interest.

\vspace{0.5cm}

\noindent \textbf{Ethical approval}

\noindent The Local Ethics Committee approved this study.

\vspace{0.5cm}

\noindent \textbf{Human and animal rights}

\noindent This article does not contain any studies with human or animal subjects performed by any of the authors.

\vspace{0.5cm}

\noindent \textbf{Consent to participate}

\noindent All participants in this study provided informed consent.

\vspace{0.5cm}

\noindent \textbf{Consent for publication}

\noindent Not required.

\section*{Acknowledgement}

\noindent This version of the article has been accepted for publication, after peer review and is subject to Springer Nature’s \href{https://www.springernature.com/gp/open-research/policies/accepted-manuscript-terms}{AM terms of use}, but is not the Version of Record and does not reflect post-acceptance improvements, or any corrections. The Version of Record is available online at: https://doi.org/10.1007/s10029-023-02939-0.

\sloppy
\printbibliography

\end{document}